\documentclass[submitted,copyright,creativecommons]{eptcs}
\providecommand{\event}{TARK 2025} 

\usepackage{iftex}
\usepackage{amsmath}

\usepackage{amsthm}
\usepackage{amsfonts}
\usepackage{amssymb}
\usepackage{graphicx}
\usepackage{color}
\usepackage{url}
\usepackage{bm}

\ifpdf
  \usepackage{underscore}         
  \usepackage[T1]{fontenc}        
\else
  \usepackage{breakurl}           
\fi

\DeclareFontFamily{U}{mathx}{}
\DeclareFontShape{U}{mathx}{m}{n}{<-> mathx10}{}
\DeclareSymbolFont{mathx}{U}{mathx}{m}{n}
\DeclareMathAccent{\widehat}{0}{mathx}{"70}
\DeclareMathAccent{\widecheck}{0}{mathx}{"71}

\title{Mechanism Design under Unawareness \\ - Extended Abstract -}
\author{Kym Pram
\institute{University of Nevada, Reno}
\email{kpram@unr.edu}
\and Burkhard C. Schipper
\institute{University of California, Davis}
\email{bcschipper@ucdavis.edu}}

\newcommand{\titlerunning}{Mechanism Design under Unawareness}
\newcommand{\authorrunning}{Kym Pram \& Burkhard C. Schipper}

\hypersetup{
  bookmarksnumbered,
  pdftitle    = {\titlerunning},
  pdfauthor   = {\authorrunning},
  pdfsubject  = {EPTCS},               
}

\newtheorem{theorem}{Theorem}

\newtheorem{proposition}{Proposition}

\newtheorem{definition}{Definition}

\begin{document}

\maketitle

\begin{abstract}
We study the design of mechanisms under asymmetric awareness and information. While the mechanism designer cannot necessarily commit to a particular social choice function in the face of unawareness, she can at least commit to properties of social choice functions such as efficiency given ex post awareness. Assuming quasi-linear utilities and private values, we show that we can implement in conditional dominant strategies a social choice function that is utilitarian ex post efficient under pooled awareness without the need of the social planner being fully aware ex ante. To this end, we develop novel dynamic versions of Vickrey-Clarke-Groves mechanisms in which true types are revealed and subsequently elaborated at endogenous higher awareness levels. We explore how asymmetric awareness affects budget balance and participation constraints. We show that ex ante unforeseen contingencies are no excuse for deficits. Finally, we propose a dynamic elaboration reverse second price auction for efficient procurement of complex incompletely specified projects with budget balance and participation constraints. 
\end{abstract}

\section{Introduction}\label{s1}

Recent years witnessed the development of logic, game theory, and decision theory with unawareness. Many contributions to the literature have been presented at prior TARK conferences (\cite{AgotnesAlechina2014, BS2024, BYG2021, DevanurFortnow2009, DFVW2018, Filiz-Ozbay2012, FritzLederman2015, Galanis2011, Galanis2013, HalpernRego2008, HalpernRego2013, HalpernRego2014, HMS2006, HMS2013a, HMS2013b, HMS2021, Heinsalu2012, Konolige1986, Ozbay2008, PVY2023, S2021, S2024}). Models of unawareness have been applied to disclosure games, moral hazard, contract theory, screening, delegation, speculation, financial market microstructure, default in general equilibrium, electoral campaigning, business strategy, and conflict resolution; for a bibliography, see \cite{S2025}. Yet, what is missing from the literature is the social engineering perspective: How to design mechanisms in the presence of unawareness? Mechanism design studies the design of institutions governing collective decisions such as markets, contracts, or political systems. Rather than defining a game and deriving its solutions like in game theory, mechanism design first identifies desirable outcomes and then designs a game such that players implement the outcome in a solution to the game. Typically, mechanisms apply to contexts with asymmetric information. However, agents may not just face asymmetric information but also asymmetric awareness. Unawareness refers to the lack of conception rather than the lack of information. Agents and the designer of mechanisms may be unaware of some events and actions affecting values and costs of complex private or public projects, allocations, and outcomes and may not even realize this fact. In this paper, we extend mechanism design to unawareness. Under the economically relevant assumption of quasi-linear preferences, we show how to design efficient mechanisms in the presence of asymmetric awareness. To this end, we introduce dynamic direct elaboration mechanisms in which not only are types communicated from agents but also awareness is raised among participants in back-and-forth communication between participants and transfers are inspired by Vickrey-Clarke-Groves (VCG) mechanisms. 

Besides our theoretical motivation for extending mechanism design to unawareness, we are motivated by practical concerns about the usefulness of explicit mechanisms in reality. For instance, it has been argued that auctions are inappropriate when projects are complex, incompletely designed, and custom-made such as the procurement of new fighter jets, buildings, consulting services, and IT projects. It is claimed that auctions may stifle communication between buyers and sellers, preventing buyers and sellers to use each other's expertise when designing projects (\cite{Goldberg1977, BajariTadelis2001, BajariMcMillanTadelis2008}). The Department of Defense (DoD), the General Service Administration (GSA), and the National Aeronautics and Space Administration (NASA) recently proposed to amend the Federal Acquisition Regulation (FAR) suggesting the prohibition of the use of reverse auctions for complex, specialized, or substantial design and construction services (\cite{OFR2024}).\footnote{When analyzing the use of auctions for procurement at the DoD, \cite{AlperBoning2003} state that auctions work best for well-specified and off-the-shelf commodities but nevertheless remark that the Navy was able to achieve substantial savings with procurement auctions for customized and complex contracts such as CVN camels, i.e., devices for safe separation of ships and piers.} The received view is that for complex projects, negotiations outperform auctions because they can better take advantage of the expertise and know-how of contractors (\cite{Goldberg1977, Sweet1994, BajariMcMillanTadelis2008}). We seek an extension of mechanism design to unawareness to make mechanisms also applicable to complex and incompletely designed projects whose efficient implementation require the pooling of expertise among participants. Our proposed mechanisms combine features of common business practices such as Request for Information (RFI) and Request for Proposals (RFP)\footnote{E.g., Federal Acquisitions Regulation 15.203.} with standard VCG mechanisms such as second price auctions. That is, we combine features of negotiations and traditional mechanism design. While traditionally in economics, negotiations have been interpreted mostly as bargaining over the surplus and thus as a substitute to mechanisms, we view negotiations more as interactively defining the surplus using the expertise of participants, which is complementary to traditional mechanisms implementing and allocating the surplus. 

Compared to typical TARK papers, this paper focuses on foundational aspects of mechanism design rather than epistemic foundations. It serves as a proof of concept that models of awareness can be fruitfully applied to fundamental problems of economic design. Moreover, our results are very general and robustly apply to no matter how beliefs and awareness are distributed among agents. From an epistemic perspective, we provide answers to questions about how awareness revelation can be incentivized among agents and how awareness can be pooled among agents via public announcements by a mediator.

This is just an extended abstract; the complete paper (\cite{PS2025}) with additional details, further results, and all proofs is available under \url{https://faculty.econ.ucdavis.edu/faculty/schipper/mechunaw.pdf} .

\section{Model\label{sec:model}} 

\subsection{Payoff Types with Unawareness\label{sec:payofftypes}} 

Let $L$ be a finite lattice with order $\trianglerighteq$. Elements of the lattice represent awareness levels. The join of the lattice is denoted by $\bar{\ell} \in L$. Define the sublattice $L(\ell) := \{\ell' \in L : \ell' \trianglelefteq \ell\}$ for $\ell \in L$. The significance of $L(\ell)$ is that an agent with awareness level $\ell$ can only reason about awareness levels in $L(\ell)$.

Fix a nonempty finite set of agents $I$. For each agent $i \in I$, there a collection of nonempty disjoint payoff type spaces $\{T_i^{\ell}\}_{\ell \in L}$. Let $\mathcal{T}_i := \bigcup_{\ell \in L} T_i^{\ell}$. A payoff type $t_i \in \mathcal{T}_i$ is more than just a value for an object. If $t_i \in T_i^{\ell}$, then describing the payoff type also requires at least awareness level $\ell$. We illustrate this feature with the following example.\\

\noindent \textbf{Example 1} Consider the context of procurement. A principal may invite agents to bid on a complex project. Agents do their due diligence and identify relevant items that drive their costs. Suppose there are items $\{a, b, c\}$. Agents may be unaware of some items even after their due diligence. Agent 1 may only be aware of items $\ell_1 = \{a, b\}$ while agent 2 is only aware of items $\ell_2 = \{b, c\}$. In this case, $L$ is isomorphic to the set of all subsets in $2^{\{a, b, c\}}$ and the natural lattice order is induced by set inclusion on $2^{\{a, b, c\}}$. Agent 1 may submit a bid given by the following table $t_1$ while agent 2 may submit a bid given by table $t_2$. 
\footnotesize
$$\begin{array}{ccc} t_1 = \begin{array}{|c|c|} \hline \mbox{Item} & \mbox{Cost} \\ \hline
\mbox{a} & 23 \\
\mbox{b} & 41 \\ \hline 
\mbox{Total} & 64 \\ \hline
\end{array} & \quad \quad \quad \quad &
t_2 = \begin{array}{|c|c|} \hline \mbox{Item} & \mbox{Cost} \\ \hline
\mbox{b} & 38 \\
\mbox{c} & 29 \\ \hline 
\mbox{Total} & 67 \\ \hline
\end{array}\end{array}$$ \normalsize As the example demonstrates, payoff types can be multi-dimensional with varying dimensions. Needless to say, the lattice approach is more general than multi-dimensional payoff type spaces.\hfill $\Box$\\

For any agent $i \in I$ and awareness levels $\ell, k \in L$ with $k \trianglerighteq \ell$, we require a surjective projection $r^{k}_{\ell}: T_i^k \longrightarrow T_i^{\ell}$ such that for all $\ell, \ell', \ell'' \in L$, $\ell'' \trianglerighteq \ell' \trianglerighteq \ell$, $r^{\ell'}_{\ell}(r^{\ell''}_{\ell'}(t_i)) = r^{\ell''}_{\ell}(t_i)$ and $r^{\ell}_{\ell}(t_i) = t_i$. For brevity, we do not index $r^k_{\ell}$ by agents. The projection relates payoff types across awareness levels. Before we illustrate this notion, we also define extensions of payoff types to greater awareness levels. For any agent $i \in I$, awareness level $\ell \in L$, and payoff type $t_i \in T_i^\ell$, we let $t_i^{\uparrow} := \bigcup_{k \trianglerighteq \ell, k \in L} (r_{\ell}^k)^{-1}(t_i)$. That is, $t_i^{\uparrow}$ is the union of inverse images at (weakly) greater awareness levels of payoff type $t_i$. Similarly, for any subset of payoff types in a given payoff type space, we let superscript ``$\uparrow$'' indicate the union of inverse images in payoff type spaces corresponding to greater awareness levels. We illustrate these notions in our prior example.\\ 

\noindent \textbf{Example 1 (Continuation)} Continuing our prior example, consider payoff types $t_1'$ and $t_1''$ of agent 1: 
\footnotesize$$\begin{array}{ccc} t_1' = \begin{array}{|c|c|} \hline \mbox{Item} & \mbox{Cost} \\ \hline
\mbox{a} & 23 \\
\mbox{b} & 41 \\ 
\mbox{c} & 16 \\ \hline 
\mbox{Total} & 80 \\ \hline
\end{array} & \quad \quad \quad \quad &  
t_1'' = \begin{array}{|c|c|} \hline \mbox{Item} & \mbox{Cost} \\ \hline
\mbox{a} & 23 \\
\mbox{b} & 41 \\ 
\mbox{c} & 15 \\ \hline 
\mbox{Total} & 79 \\ \hline
\end{array} \end{array}$$ \normalsize Both payoff types are described with items in $\{a, b, c\}$. That is, both $t_1', t_1'' \in T_1^{\{a, b, c\}}$. When payoff information on item $c$ is stripped away from payoff type $t_1'$ we obtain payoff type $t_1$. That is, $r^{\{a, b, c\}}_{\{a, b\}}(t_1') = t_1$. Similarly, $r^{\{a, b, c\}}_{\{a, b\}}(t_1'') = t_1$. Thus, $t_1, t_1', t_1'' \in t_1^{\uparrow}$. \hfill $\Box$\\

For any agent $i \in I$, we define $\lambda: \mathcal{T}_i \longrightarrow L$ by $\lambda(t_i) = \ell$ if $t_i \in T_i^{\ell}$. Again, for brevity we do not index $\lambda$ by agents. The function $\lambda$ indicates the awareness level that is required to describe the payoff type. For instance, in Example 1, $\lambda(t_1) = \{a, b\}$ while $\lambda(t_1') = \{a, b, c\}$. 

Our framework is general enough to capture payoff types described by all kinds of formal objects like sets of formulae in a formal language (e.g., \cite{HMS2008}), abstract sets, vectors, matrices, formal concepts, pre-sheaves etc. that may represent verbal, quantitative, or pictorial features of proposals, tenders, quotations, messages etc. in business practice. By abstracting from these particular features, we obtain a theory that works for more than one kind of formalism, ensure tractability, and focus on what is really essential for modeling awareness levels, namely the existence of an order of expressiveness of descriptions.  

For each agent $i \in I$, payoff types and awareness levels are drawn consistently as follows: At awareness level $\bar{\ell}$, nature draws a payoff type $\bar{t}_i \in T_i^{\bar{\ell}}$ in the upmost payoff type space, interpreted as agent $i$'s true payoff type if she were aware of everything, and an awareness level $\ell_i \in L$. Consequently, the agent's perceived payoff type is $r^{\bar{\ell}}_{\ell_i}(t_i)$. That is, agent $i$ can ``miss something'' but he cannot perceive the ``wrong'' payoff type w.r.t. what he is aware. More generally, for any awareness level $\ell \in L$, nature draws agent $i$'s corresponding type $r^{\bar{\ell}}_{\ell}(\bar{t}_i)$ and agent $i$'s awareness level $\ell' = \ell_i \wedge \ell$. This means in particular, that if $\ell = \ell_i$, then it is payoff type $r^{\bar{\ell}}_{\ell_i}(t_i)$ and awareness level $\ell_i$ as just discussed. If $\bar{\ell} \trianglerighteq \ell \trianglerighteq \ell_i$, nature draws agent $i$'s corresponding type $r^{\bar{\ell}}_{\ell}(\bar{t}_i)$ and awareness level $\ell_i$. If $\ell \in L$ with $\ell_i \trianglerighteq \ell$, nature draws agent $i$'s corresponding type $r^{\bar{\ell}}_{\ell}(\bar{t}_i)$ and awareness level $\ell$. Finally, if $\ell \in L$ with $\ell \not\trianglerighteq \ell_i$, nature draws agent $i$'s corresponding type $r^{\bar{\ell}}_{\ell}(\bar{t}_i)$ and awareness level $\ell' = \ell_i \wedge \ell$. This specifies perceived payoff types and awareness levels consistently across the type spaces $\{T_i^{\ell}\}_{\ell \in L}$. Notice that if $t_i \in T_i^{\ell}$ for some $\ell \in L$, then the agent's perceived type is always in a type space with (weakly) less awareness than $\ell$. 
 
For any $\ell \in L$, let $\bm{T}^{\ell} := \times_{i \in I} T_i^{\ell}$. Moreover, let $\boldsymbol{\mathcal{T}} := \times_{i \in I} \mathcal{T}_i$. Similarly, we let $\bm{T}^{\ell}_{-i} := \times_{j \in I \setminus \{i\}} T_j^{\ell}$ and $\boldsymbol{\mathcal{T}}_{-i} := \times_{j \in  \in I \setminus \{i\}} \mathcal{T}_j$. 

When agents have the payoff type profile $\bm{t} = (t_i)_{i \in I} \in \boldsymbol{\mathcal{T}}$, their \emph{pooled} awareness level is $\widecheck{\lambda}(\bm{t}) := \bigvee_{i \in I} \lambda(t_i) \in L$, i.e., the join of all agent's awareness levels at payoff type profile $\bm{t}$. Since $L$ is a finite lattice, the join always exists in $L$. Note that $\widecheck{\lambda}(\bm{t})$ may be a greater awareness level than any of the agent's awareness levels. 

For each $\ell \in L$, there is a nonempty compact set of outcomes or allocations $X^{\ell}$. This formulation allows for unawareness of outcomes. We require that $\ell \trianglerighteq k$ implies $X^k \subseteq X^{\ell}$. Denote by $X := \bigcup_{\ell \in L} X^{\ell}$. Clearly, $X = X^{\bar{\ell}}$.

Each agent $i \in I$ has an upper semi-continuous utility function $u_i: \bigcup_{\ell \in L} X^\ell \times T_i^\ell  \longrightarrow \mathbb{R}$. From this formulation it is clear that we focus on private payoff types. To facilitate a commonly used notion of efficiency, we assume that each agent's utility function is quasilinear. I.e., for each $i \in I$, $u_i(x, t_i) := v_i(x_0, t_i) + x_i$ for $x = (x_0, x_1, ..., x_{|I|}) \in X^\ell := X_0^\ell \times \mathbb{R}^{|I|}$ and $t_i \in T_i^\ell$, $\ell \in L$. As usual, $x_0$ describes the physical properties of the outcome while $(x_i)_{i \in I}$ represents the vector of transfers made \emph{to} agents.

We denote the outcome function by $f_0: \boldsymbol{\mathcal{T}} \longrightarrow X_0$. That is, $f_0(\bm{t})$ is the physical outcome prescribed by the outcome function $f_0$ to type profile $\bm{t}$. We require that for any $\bm{t} \in \boldsymbol{\mathcal{T}}$, $f_0(\bm{t}) \in X_0^{\widecheck{\lambda}(\bm{t})}$. That is, the social planner can use joint awareness to select the outcome/allocation. If the reported payoff type profile is $\bm{t}$, then the planner's awareness is $\widecheck{\lambda}(\bm{t})$ and hence outcomes in $X_0^{\widecheck{\lambda}(\bm{t})}$ are selected.\footnote{Typically the planer is not considered as an agent in mechanism design and hence we do not explicitly consider the awareness that the planner may have. Yet, all of our results remain intact when the planner joins her awareness together with the awareness of all agents whenever communicating the pooled awareness level in the dynamic elaboration mechanisms introduced in the next section.} The formulation also makes clear that the social planner cannot necessarily describe the outcome function to agents in advance if the social planner is unaware of something herself. However, we assume that the social planner can commit to abstract properties of outcome functions such as efficiency, which in a quasi-linear setting takes the following form: 

\begin{definition}\label{def:efficient} The outcome function $f_0$ is \emph{utilitarian ex-post efficient} if for all $\ell \in L$ and $\bm{t} = (t_i)_{i \in I} \in \bm{T}^\ell$,
\begin{align} \sum_{i \in I} v_i(f_0(\bm{t}), t_i) & \geq \sum_{i \in I} v_i(x_0, t_i) \mbox{ for all } x_0 \in X_0^{\ell}.
\end{align}
\end{definition}

\subsection{Dynamic Direct Elaboration Mechanisms\label{sec:DDEM}} 

In the appendix, we show that VCG mechanisms are insufficient for efficient implementation under unawareness as they fail to pool awareness. In order to pool awareness among agents and allow agents to provide information on issues they are or became aware, we allow for communication not just \emph{from} agents but also \emph{to} agents. That is, we envision a mediator who receives messages about payoff types from agents, pools the awareness contained in those messages, and sends back messages to agents that potentially raise the awareness of agents to the pooled awareness level. Subsequently, agents may want to elaborate on their prior messages at least to the details of the pooled awareness level. This procedure may be repeated till no agent wants to elaborate further. To this end, we introduce a new class of dynamic mechanisms. 

\begin{definition}[Dynamic Direct Elaboration Mechanism]\label{def:directelaborationmechanism} The \emph{dynamic direct elaboration mechanism} implementing outcome function $f_0$ is defined recursively by the following algorithm:\footnote{Transfers can be arbitrary at this point and will be specified later when we consider particular dynamic direct elaboration mechanisms.}
\begin{itemize}
\item[] Stage $n = 1$: Each agent $i \in I$ must report a type $t_i^1$.

\item[] Stages $n > 1$: Each agent $i$ must report a type $t_i^n \in (t_i^{n-1})^{\uparrow} \cap \left(T_i^{\widecheck{\lambda}(\bm{t}^{n-1})}\right)^{\uparrow}$. 

\item[] Stop: If $t_i^{n+1} = t_i^{n}$ for all $i \in I$, then $f_0(\bm{t}^{n+1})$ is implemented.

\end{itemize}
\end{definition} 

At the first stage, every agent reports a type. At later stages, agents can elaborate on their prior reported types. When agents report the payoff type profile $\bm{t}^{n-1}$ in stage $n - 1$, the mediator awareness level is the pooled awareness level $\widecheck{\lambda}(\bm{t}^{n-1})$. Consequently, he communicates back the pooled awareness level $\widecheck{\lambda}(\bm{t}^{n-1})$ to all agents.\footnote{The current formulation presumes that the mediator, mechanism designer, or social planner has no relevant awareness herself. If she does have relevant awareness, she can incorporate it easily into the message communicated back to agents.} Importantly, all agents receive the same message from the mediator. That is, we can use public messages. At the next stage, each agent $i$ must report a type at least at the pooled awareness level  $T_i^{\widecheck{\lambda}(\bm{t}^{n-1})}$. This report must be consistent with her prior reported type which is implied by the requirement $(t_i^{n-1})^{\uparrow} \cap \left(T_i^{\widecheck{\lambda}(\bm{t}^{n-1})}\right)^{\uparrow}$. For instance, if an agent reported a type whose awareness is strictly below the pooled awareness of all agents' reports, then she must now report a type at least at the pooled awareness level $\widecheck{\lambda}(t_i^{n-1})$ that is consistent with her prior reported type, i.e., a type in $(t_i^{n-1})^{\uparrow}$. However, it could be the case that she has previously pretended to have much less awareness so that the pooled awareness level did not yet incorporate all her awareness. Thus, we allow her to report a type with awareness strictly larger than the pooled awareness level. This motivates the requirement $t_i^n \in (t_i^{n-1})^{\uparrow} \cap \left(T_i^{\widecheck{\lambda}(\bm{t}^{n-1})}\right)^{\uparrow}$ rather than $t_i^n \in (t_i^{n-1})^{\uparrow} \cap T_i^{\widecheck{\lambda}(\bm{t}^{n-1})}$. If an agent did report at the pooled awareness level in the previous stage, then she is not able to change her report in the current stage. The mechanism stops once no agents revise their reports any further. Once it stops, it implements the physical outcome associated to the final reported type profile by $f_0$. \\

\noindent \textbf{Example 1 (Continued)} To continue our Example 1 above, suppose that agent 1 reported in stage 1 payoff type $t_1^1 = t_1$ and agent 2 reported $t_2^1 = t_2$. Pooling awareness leads to $\widecheck{\lambda}(t_1, t_2) = \{a, b, c\}$. At stage 2, both agents must now elaborate their prior reported type and report a type in $T_i^{\{a, b, c\}}$. For instance, agent 1 could report $t_1^2 = t_1'$. Since no further awareness can be revealed after stage 2, the mechanism must conclude after stage 3. \hfill $\Box$\\

Note that awareness and information is transmitted both from agents to the mediator but also from the mediator to the agents. Thus, awareness of agents may change endogenously when interacting in the mechanism. Since agents report directly types and elaborate in later stages one their prior reported types, we call it a ``direct elaboration'' mechanism. Note that by the definition of join of the lattice of spaces, $\widecheck{\lambda}(\bm{t}^{n-1}) \trianglerighteq \lambda(t_i^{n-1})$ for any $i \in I$.  

The mechanism stops when no agent wants to further elaborate on her type. Clearly, since $L$ is finite, the mechanism must stop at some finite stage $n$. Moreover, when it stops at $n$, then $\bm{t}^n \in \bm{T}^{\widecheck{\lambda}(\bm{t}^n)}$. That is, all agents must have reported twice in a row types at the \emph{same} awareness level.  

Denote by $\sigma_i$ agent $i$'s reporting strategy, by $\sigma_i^*$ agent $i$'s truth-telling strategy, by $\bm{\sigma}$ a strategy profile, by $h_i$ an information set of agent $i$'s, $\Sigma_i(h_i)$ agent $i$'s strategies that allow information set $h_i$, $\mathfrak{T}$ the set of sequences of reported payoff type profiles, by $\tau(\bm{t}, \bm{\ell}, \bm{\sigma})$ the profile of reported payoff type profiles the final when $\bm{t}$ is the profile of payoff types, $\bm{\ell}$ is the profile of awareness levels, and $\bm{\sigma}$ is the profile of strategies, by $\tau^*(\bm{t}, \bm{\ell}, \bm{\sigma})$ the final reported payoff type profile thereof. We also extend definition of $\lambda$ to information sets. In the complete manuscript (\cite{PS2025}), we provide details on the dynamic game with unawareness induced by dynamic direct elaboration mechanisms, information sets, and strategies. 

For any agent $i \in I$, let $f_i: \mathfrak{T} \longrightarrow \mathbb{R}$ denote the transfer paid \emph{to} agent $i$ in the dynamic direct elaboration mechanisms. These transfers will depend on the precise versions of the direct elaboration mechanism studied below. In contrast to the outcome function $f_0$ we allow transfers to depend on the entire sequence of reported type profiles.

We let the social choice function (i.e., the outcome function and transfers) be denoted by $f: \mathfrak{T} \longrightarrow \mathbb{R}$ and defined by $f = (f_0, f_1, ..., f_{|I|})$. That is, for any strategy profile $\bm{\sigma} \in \bm{\Sigma}$, $\ell \in L$, initial profiles of payoff types $\bm{t} \in \bm{T}^{\ell}$, initial profiles of awareness levels $\bm{\ell} \in L(\ell)^{|I|}$, $f(\tau(\bm{t}, \bm{\ell}, \bm{\sigma})) = \left(f_0(\tau^*(\bm{t}, \bm{\ell}, \bm{\sigma})), f_1(\tau(\bm{t}, \bm{\ell}, \bm{\sigma})), ...,\right.$ $\left.f_{|I|}(\tau(\bm{t}, \bm{\ell}, \bm{\sigma}))\right)$.  

\begin{definition}\label{defin:conditionaldominance} The dynamic direct elaboration mechanism truthfully implements the social choice function $f$ in conditional dominant strategies if for all agents $i \in I$, information sets $h_i \in H_i(\sigma^*_i)$, opponents' strategy profiles $\bm{\sigma}_{-i} \in \bm{\Sigma}_{-i}$, initial profiles of payoff types $\bm{t} \in \bm{T}^{\lambda(h_i)}$ (as perceived by $i$ in $h_i$), and initial profiles of awareness levels $\bm{\ell} \in L(\lambda(h_i))^{|I|}$ (as perceived by $i$ in $h_i$) such that $h_i \in H_i(\bm{t}, \bm{\ell}, (\sigma^*_i, \bm{\sigma}_{-i}))$, 
\begin{eqnarray} u_i(f(\tau(\bm{t}, \bm{\ell}, (\sigma_i^*, \bm{\sigma}_{-i})), t_i(h_i))  & \geq & u_i(f(\tau(\bm{t}, \bm{\ell}, (\sigma_i, \bm{\sigma}_{-i})), t_i(h_i)) \label{eqn:conditionaldominance} 
\end{eqnarray} for all $\sigma_i \in \Sigma_i(h_i)$. 
\end{definition}

Conditional dominance strengthens dominance by requiring each agent not only to select a (partial) strategy that is ex-ante dominant but also dominant conditional on each information set. This becomes important when agents cannot anticipate all information sets ex-ante and thus cannot select ex-ante a strategy for the entire game in extensive form. The use of conditional dominance as a solution concept for implementing outcomes in our mechanisms implies that our results do not depend on agents' beliefs about other agents, their awareness etc. That is, revelation of awareness and private information in the mechanisms is robust to possibly misspecified beliefs by the mediator and agents.  

\begin{definition}\label{def:pooledawareness} A outcome function $f_0$ is truthfully implemented at the pooled awareness level if for any $\bm{\bar{t}} \in \bm{T^{\bar{\ell}}}$ and $(\ell_i)_{i \in I} \in L^{|I|}$, $f_0\left(r^{\bar{\ell}}_{\left(\bigvee_{i \in I} \ell_i\right)} (\bm{\bar{t}})\right)$ is implemented. 
\end{definition} 

Recall that $\left(\bm{\bar{t}}, (\ell_i)_{i \in I}\right) \in \bm{T^{\bar{\ell}}} \times L^{|I|}$ represents the move of nature selecting both actual payoff types and awareness levels for all agents. Consequently, $\bigvee_{i \in I} \ell_i$ represents the pooled awareness level.

\section{Efficient Implementation\label{sec:DEVCGM}} 

To implement efficiently at the pooled awareness level, we specify for each agent $i \in I$ the transfers $f_i$ that incentivize both the revelation of information and raising awareness. Revelation of information is achieved via VCG transfers. Raising awareness requires an extra term in the transfer functions. 

\begin{definition}[Dynamic Elaboration VCG Mechanism]\label{def:VCGmechanism} We say that the dynamic direct elaboration mechanism implementing $f$ is a dynamic elaboration VCG mechanism if $f_0$ is utilitarian ex-post efficient and transfers $f_i$ to agent $i \in I$ are given by for any $(\bm{t}^1, ..., \bm{t}^n) \in \mathfrak{T}$,
\begin{eqnarray}\label{eqn:amendedVCGtransfers1} f_i(\bm{t}^1, ..., \bm{t}^n) := \sum_{j \neq i} v_j(f_0(\bm{t}^n), t_j^n) + y_i^{\widecheck{\lambda}(\bm{t}^n)}(\bm{t}^n_{-i}) + a_i(\bm{t}^1, ..., \bm{t}^n),
\end{eqnarray}
where, for each $\ell$, $y_i^{\ell}: \bm{T}^{\ell}_{-i} \longrightarrow \mathbb{R}$ is an arbitrary function and $a_i(\cdot)$ is defined by:
\begin{eqnarray}\label{eqn:amendedVCGtransfers2}
a_i(\bm{t}^1, ..., \bm{t}^n) :=
\begin{cases}
m_i(\widecheck{\lambda}(\bm{t}^n)) & \mbox{if } i = i^*(\bm{t}^1, ..., \bm{t}^n) \\
-\frac{1}{|I|-1} m_j(\widecheck{\lambda}(\bm{t}^n)) & \mbox{if } j = i^*(\bm{t}^1, ..., \bm{t}^n) \neq i \\
0 & \mbox{otherwise}
\end{cases}
\end{eqnarray}
where 
\begin{eqnarray*} i^*(\bm{t}^1, ..., \bm{t}^n) & := & \left\{i \in I : \lambda(t_i^k) = \widecheck{\lambda}(\bm{t}^n) \mbox{ and } \not\exists j \neq i, k' \leq k \mbox{ such that } \lambda(t_j^{k'}) = \widecheck{\lambda}(\bm{t}^n)\right\}
\end{eqnarray*} and $m_i(\ell)$ is defined recursively, as follows: $m_i(\underline{\ell}) :=  0$ and for any $\ell \triangleright \underline{\ell}$,
\begin{eqnarray}\label{eqn:amendedVCGtransfers3} m_i(\ell)  & := &  \max\limits_{\substack{\ell' \triangleleft \ell \\ \bm{t}' \in \bm{T}^{\ell'} \\ \bm{t} \in \bm{T}^{\ell}}} \left( m_i(\ell') + v_i(f_0(\bm{t}'), t_i) + \sum_{j\neq i} v_j(f_0(\bm{t}'), t'_j) + y_i^{\ell'}(\bm{t}'_{-i}) - \sum_{j} v_j(f_0(\bm{t}), t_j) - y_i^\ell(\bm{t}_{-i})\right).
\end{eqnarray} 
\end{definition}

The mechanism can viewed as a dynamic version of the Vickrey-Clarke-Groves (VCG) mechanisms (\cite{Groves1973, GrovesLoeb1975}). Note that the transfers can be described without necessarily being aware of all payoff type profiles. The mechanism designer commits to implement a utilitarian ex-post efficient outcome given the agents' final reports of payoff types. Each agent is paid the total welfare of others given the final reports of payoff types plus a term that depends on only the opponents' payoff types (and the final pooled awareness level) and additionally a term incentivizing raising of awareness. This last term does not only depend on the final reported payoff type profile but on the sequence of reported payoff type profiles. It matters who reports the pooled awareness level first. Note that $i^*(\bm{t}^1, ..., \bm{t}^n) = \emptyset$ if there is no agent who reports the pooled awareness level (which can happen if the individual awareness levels are incomparable) or if several agents simultaneously report the pooled awareness level first. If $i$ is the unique agent who first reports the pooled awareness level, then $a_i$ receives a payment related to the cost that raising awareness could impose on her utility from the mechanism, net of the $a_i(\cdot)$ term itself. 

\begin{theorem}\label{theo:VCG} The dynamic elaboration VCG mechanism truthfully implements in conditionally dominant strategies a utilitarian ex-post efficient outcome under pooled awareness.  
\end{theorem}

The proof is contained the complete manuscript (\cite{PS2025}). 

\begin{proposition} The utilitarian ex-post efficient outcome under pooled awareness is truth-fully implemented in conditional dominant strategies in the game induced by the dynamic elaboration VCG mechanism in at most three stages.
\end{proposition}

\section{No Deficit\label{sec:ND}} 

Ideally, we like our mechanisms to satisfy desirable properties beyond efficiency. In the complete manuscript, we characterize budget balance of the dynamic elaboration VCG mechanism and show that it does not impose constraints beyond static VCG mechanisms. In classical mechanism design, it is well known that the VCG mechanisms like the Groves mechanisms can not satisfy budget balance in general (\cite{GreenLaffont1979}). We may want to look for weaker requirements. Not being budget balanced means that the mechanisms can run a deficit or surplus. In the context of unawareness, we are interested more in running no deficit than running no surplus for two reasons: First, if the mechanisms runs a deficit larger than the mechanism designer anticipated, agents may suspect that the mechanism designer is not committed to the mechanisms and become reluctant to truth-fully report their payoff types. Second, unforeseen contingencies are often used to justify budget overruns. But are deficits really inevitable in the presence of unawareness? 

Recall that we used the convention that transfers $f_i$ denote transfers \emph{to} agent $i$. The following property is sometimes also called weak budget balance (e.g., \cite{ShohamLeyton-Brown2012}). 

\begin{definition}[No deficit]\label{def:nodeficit} We say that the dynamic direct elaboration mechanism with transfer functions $(f_i)_{i \in I}$ satisfies no deficit if for all $(\bm{t}^1, ..., \bm{t}^n) \in \mathfrak{T}$, 
\begin{eqnarray} \sum_{i \in I} f_i(\bm{t}^1, ..., \bm{t}^n) & \leq & 0. \label{eqn:notransfers}
\end{eqnarray}
\end{definition} 

In classical mechanism design, it is well known that a Groves mechanism may run a deficit, but the Clarke (or Pivot) mechanism, in which each agent pays the negative externality that they impose on other agents through their effect on the social choice, does not. Therefore, the idea is to design a version of dynamic elaboration VCG mechanisms with Clarke transfers and show that it does not run a deficit. 

To define the mechanism, we have to specify externalities. For any agent $i \in I$, define the $(-i)$-utilitarian ex-post efficient outcome function $f^{-i}_0: \boldsymbol{\mathcal{T}} \longrightarrow X_{0}$ by $f^{-i}_0(\bm{t}) = \arg \max_{x_{0} \in X^{\widecheck{\lambda}(\bm{t})}_{0}} \sum_{j \neq i} v_j(x_{0}, t_j)$. That is, for each profile of payoff types $\bm{t} \in \boldsymbol{\mathcal{T}}$, $f^{-i}_0$ maximizes utilitarian ex-post welfare taking into account only the value functions of agent $i$'s opponents. Note that different from restricted outcome functions in standard Clarke mechanism the argument of $f^{-i}_0$ is the \emph{full} profile of payoff types $\bm{t} = (t_j)_{j \in I}$ rather than just $\bm{t}_{-i}$. The reason is that in order to compute $f^{-i}_0$ we need the awareness of \emph{all} agents, not just agents $j \neq i$: Although agent $i$ value is not considered when evaluating the social welfare of an outcome, for agents $j \neq i$, the efficiency of this outcome is still evaluated at the pooled awareness level.

\begin{definition}[Dynamic Elaboration Clarke Mechanism]\label{def:clarkemechanism} We say that the dynamic direct elaboration mechanism implementing $f$ is a dynamic elaboration Clarke mechanism if it is a dynamic elaboration VCG mechanisms with, for all $\bm{t}^n \in \bm{T}^{\widecheck{\lambda}(\bm{t}^n)}$ and $i \in I$, 
$$y_i^{\widecheck{\lambda}(\bm{t}^n)}(\bm{t}^n_{-i}) := - \sum_{j \neq i} v_j(f^{-i}_0(\bm{t}^n), t^n_j).$$
\end{definition}

Observe that for any $i \in I$, if $\sum_{j \neq i} v_j(f^{-i}_0(\bm{t}^n), t^n_j) \geq \sum_{j \neq i} v_j(f_0(\bm{t}^n), t_j^n)$ then $f_i(\bm{t}^1, ..., \bm{t}^n) - a_i(\bm{t}^1, ..., \bm{t}^n)$ is non-positive. Since $f^{-i}_0(\bm{t}^n)$ maximizes the sum of values over $j \in I \setminus \{i\}$, we have $\sum_{j \neq i} v_j(f^{-i}_0(\bm{t}^n), t^n_j) \geq \sum_{j \neq i} v_j(f_0(\bm{t}^n), t_j^n)$. Together with budget neutrality of the $a_i$-terms, this implies now that the mechanism satisfies no deficit. Utilitarian ex-post efficiency is implied by Theorem~\ref{theo:VCG}. 

\begin{theorem}\label{theo:Clarke} The dynamic elaboration Clarke mechanism truthfully implements in conditionally dominant strategies a utilitarian ex-post efficient outcome under pooled awareness with no deficit.  
\end{theorem}

The proof is contained the complete manuscript (\cite{PS2025}).

We illustrate the dynamic elaboration Clarke mechanisms with two examples. In the first example, neither agent will announce the pooled awareness level as the pooled awareness level is the join that is strictly greater than any agent's awareness.\\ 

\noindent \textbf{Example 1 (Continuation)} Consider again Example 1. In the conditional dominant solution to the dynamic elaboration Clarke mechanism, agents report in the first stage respectively, 
\footnotesize $$\begin{array}{ccc} t_1 = \begin{array}{|c|c|} \hline \mbox{Item} & \mbox{Cost} \\ \hline
\mbox{a} & 23 \\
\mbox{b} & 41 \\ \hline 
\mbox{Total} & 64 \\ \hline
\end{array} & \quad \quad \quad \quad & t_2 = \begin{array}{|c|c|} \hline \mbox{Item} & \mbox{Cost} \\ \hline
\mbox{b} & 38 \\
\mbox{c} & 29 \\ \hline 
\mbox{Total} & 67 \\ \hline
\end{array}
\end{array}$$ \normalsize  
Since $\lambda(t_1) = \{a, b\}$ and $\lambda(t_2)=\{b, c\}$, agents are made aware of the join $\widecheck{\lambda}(t_1, t_2) = \{a, b, c\}$ and are invited to report an elaborated type in $T_i^{\{a, b, c\}}$, $i \in {1, 2}$, respectively. Extending the example slightly, let agents report truthfully their elaborations, respectively, 
\footnotesize $$\begin{array}{ccc} t'_1 = \begin{array}{|c|c|} \hline \mbox{Item} & \mbox{Cost} \\ \hline
\mbox{a} & 23 \\
\mbox{b} & 41 \\ 
\mbox{c} & 16 \\ \hline 
\mbox{Total} & 80 \\ \hline
\end{array} & \quad \quad \quad \quad & t'_2 = \begin{array}{|c|c|} \hline \mbox{Item} & \mbox{Cost} \\ \hline
\mbox{a} & 19 \\
\mbox{b} & 38 \\
\mbox{c} & 29 \\ \hline 
\mbox{Total} & 86 \\ \hline
\end{array}
\end{array}$$ \normalsize 
To complete the example, suppose that the possible physical outcomes (at any awareness level) are that the good is produced by agent 1, the good is produced by agent 2 or the good is not produced, i.e., $X_0 = \{1, 2, \emptyset\}$. Finally, let there be a third agent, a buyer, who always values the good at 100 no matter whether it comes from agent 1 or agent 2, i.e., $v_3(1, t_3) = v_3(2, t_3) = 100$ and $v_3(\emptyset, t_3) = 0$ for any $t_3$. 

Then, the efficient decision at the updated type profile is for the good to be produced by agent 1 (at a total cost of 80). Since no player is the first to announce the joint awareness level, the $a_i(\cdot)$-term is $0$ for each agent. Agent 3 is pivotal in the sense that if agent 3's valuation is not considered, the good would not be produced at all. We have for all $\bm{t}$, $f(\bm{t}) = 1$, $f^{-1}(\bm{t}) = f^{-2}(\bm{t}) = 1$, and $f^{-3}(\bm{t}) = \emptyset$. The transfers to agent 1 are $100 - 0 - (100 - 0) + 0 = 0$, to agent 2 they are $20 - (100 - 80) - 0 = 0$, and to agent 3 are $-80 - 0 + 0 = -80$. Thus, the mechanism runs a surplus of $80$.\footnote{Note that the mechanism does not satisfy agent 1's ex-post participation constraints. This is true of the Clarke mechanism in this context even without any unawareness. It is known that the Clarke mechanisms may not necessarily satisfy ex-post participation constraints. We will analyze participation constraints in the next section.} \hfill $\Box$\\

Note that in this example awareness did not improve utilitarian ex-post welfare. Nevertheless, we are able to raise awareness to the pooled awareness level with out mechanisms. 

In the complete manuscript (\cite{PS2025}), we provide additional examples illustrating the novel $a_i$-terms are non-trivial.

In the complete manuscript (\cite{PS2025}), we study participation constraints and show that the dynamic elaboration Clarke mechanism satisfies ex ante anticipated ex post participation constraints under standard assumptions on the value functions. Finally, also in the complete manuscript (\cite{PS2025}) we introduce dynamic elaboration reverse second price auctions for procurement contexts and show that they implement efficient outcomes under budget balance and ex post participation constraints.

\section{Related Literature} 

Our paper contributes to the releated recent literature on contracting under unawareness (e.g., \cite{Lee2008, SommerLoch2009, vonThaddenZhao2012, Auster2013, Filiz-Ozbay2012, GrantKlineQuiggin2012, ChungFortnow2016, AusterPavoni2024, LeiZhao2021, FrancetichSchipper2024}). Closer to mechanism design, \cite{LiSchipper2024} study the seller’s decision to raise bidders’ awareness of characteristics before a second-price auction with entry fees. Optimal entry fees capture an additional unawareness rent due to unaware bidders misperceiving their probability of winning and the price to be paid upon winning. In contrast to our setting, the auctioneer is aware of everything ex-ante. \cite{Piermont2024} studies how a decision maker can incentivize an expert to reveal novel aspects about a decision problem via an iterated revelation mechanism in which at each round the expert decides on whether or not to raise awareness of a contingency that in turn the decision maker considers when proposing a new contract that the expert can accept or reject. This bears some similarity with our dynamic direct elaboration mechanisms. Most importantly, we focus on a multi-agent setting with transferable utilities. \cite{Piermont2024} shows that iterated revelation allows for efficient outcomes to emerge. 

\cite{HerwegSchmidt2020} study a procurement problem with a principal and two agents who may be aware of some design flaws. Our paper differs from theirs in many respects: First, in \cite{HerwegSchmidt2020} agents can raise awareness of realized design flaws, agent's private costs are independent of the design flaw, and fixing the design flaw requires a known common cost that is interim verified by an industry expert. In our model, agents report payoff types, can raise awareness of potential factors affecting payoffs, and individually elaborate how the payoffs change in light of new awareness. Second, \cite{HerwegSchmidt2020} construct an efficient direct mechanism under common awareness and then argue that there is an indirect mechanism that under asymmetric awareness that gives rise to the same outcomes and incentives. This leads them to conclude that there are also corresponding equilibria in these two mechanisms. Yet, an appropriate notion of equilibrium in mechanisms under asymmetric awareness should verify equilibrium behavior w.r.t what outcomes agents anticipate and how behavior is affect by changing anticipations of outcomes during the play. Our approach is more ``direct'': We conduct our entire analysis of efficient conditional dominant strategy implementation in a direct mechanisms under asymmetric awareness thereby modeling every potential deviation from equilibrium behavior from the agent's point view. Finally, \cite{HerwegSchmidt2020} focus on the particular but very relevant application to procurement while we consider more generally the efficient mechanism design problem under asymmetric awareness. 

\bigskip 

\noindent \textbf{Acknowledgment } Burkhard gratefully acknowledges financial support via ARO Contract W911NF2210282. 

\bibliographystyle{eptcs}
\bibliography{mechunaw}

\end{document}